\documentstyle[preprint,aps,prl,epsfig,amssymb]{revtex}
\tightenlines
\newcommand{\<}{\langle}
\renewcommand{\>}{\rangle}
\newcommand{\be}{\begin{equation}}
\newcommand{\ee}{\end{equation}}

\begin{document}
\title{A Monte Carlo study of the triangular lattice gas with
the first- and the second-neighbor exclusions}
\author{Wei Zhang~$^{1}$ and
Youjin Deng~$^{2}$~\footnote{yd10@nyu.edu} }
\address{$^{1}$ Department of Physics,
Ji-Nan University, Guangzhou 510630, China}
\address{$^{2}$Physikalisches Institut, Universit\"at
Heidelberg, Philosophenweg 12, 69120 Heidelberg, Germany}

\date{\today}
\maketitle
\begin{abstract}
We formulate a Swendsen-Wang-like version of the geometric cluster
algorithm. As an application,
we study the hard-core lattice gas on the triangular lattice
with the first- and the second-neighbor exclusions.
The data are analyzed by finite-size scaling, but the 
possible existence of logarithmic corrections is not
considered due to the limited data. We determine
the critical chemical potential as $\mu_c=1.75682 (2)$ and the
critical particle density as $\rho_c=0.180(4)$. 
The thermal and magnetic exponents $y_t=1.51(1) \approx 3/2$ and
$y_h=1.8748 (8) \approx 15/8$, estimated from Binder ratio $Q$ and 
susceptibility $\chi$, strongly support the general belief that
the model is in the 4-state Potts universality class.
On the other hand, the analyses of energy-like quantities yield 
the thermal exponent $y_t$ ranging from $1.440(5)$ to $1.470(5)$.
These values differ significantly from the expected value $3/2$,
and thus imply the existence of logarithmic corrections.
\end{abstract}
\section{Introduction}
Lattice gases, together with the Potts (including Ising) and the O$(n)$
model, play an important role in the statistical mechanics.
They are used to describe
universal properties of many complex physical systems, ranging
from simple fluids to structural glasses and granular materials.
Lattice-gas models are generally defined as follows. For
a given lattice, a number of particles is randomly 
distributed over its vertices with
the constraint that each vertex can at most be occupied by one
particle; the density of particles is controlled by the chemical
potential $\mu$. Particles on different vertices can interact
with one another--normally through two-body interactions. Accordingly,
the reduced Hamiltonian (already divided by $kT$ with 
Boltzmann factor $k$ and temperature $T$)
of a lattice-gas model can be written as
\begin{equation}
{\mathcal H} = - \mu \sum_{i} \sigma_i
- K_{1N} \sum_{\{jk\}}  \sigma_j \sigma_k
- K_{2N} \sum_{\{ lm \}}  \sigma_l \sigma_m  +\cdots \; ,
\label{def_H_Gas}
\end{equation}
where $\sigma=0,1$ represents the absence and the presence of a particle,
respectively. The second term with amplitude $K_{1N}$ describes the
first-neighbor interactions, and the third one with $K_{2N}$ is for the
second-neighbor couplings; further-neighbor interactions can be
included, as denoted by symbol $\cdots$.

In the study of lattice gases, one often takes the hard-core
limit: in Eq.~(\ref{def_H_Gas}) the couplings $K$ outside a certain
range $(r>r_0)$ are set zero, while $ K( r \leq r_0)$ is
taken to the limit $K \rightarrow -\infty$; namely,
the particles have a hard-core of radius $r_0$. 
A particular example is the lattice-gas model with nearest-neighbor
exclusion: $K_{1N} \rightarrow -\infty$ while all further-neighbor 
couplings are zero. Unlike the Potts model and the
O$(n)$ spin model, the nature and the universality of the
phase transitions in lattice-gas systems depend on the lattice
structures. For instance, the lattice-gas model
with nearest-neighbor exclusions on the square and the honeycomb
lattice is believed to be Ising-like, while that on the triangular lattice
(Baxter's hard-hexagon model~\cite{Baxter_80,Baxter_82})
belongs to the 3-state Potts universality class.

Extensive investigations have been carried out for lattice gases, and
many theoretical and numerical approaches are applied. This
includes exact calculations (mainly by Baxter and coauthors),
series expansions (like high-temperature and low-temperature expansions),
cluster variation method, transfer matrix calculations, and Monte Carlo
simulations etc. The critical free energy of Baxter's hard-hexagon 
lattice gas was exactly 
calculated~\cite{Baxter_80,Baxter_81,Huse_82,Baxter_83}; 
the critical chemical potential is
known as $\mu_c= \ln [(11+\sqrt{5})/2] $, and the critical particle
density is $\rho_{c}=(5+\sqrt{5})/10$.
Baxter's hard-square model~\cite{Baxter_80,Baxter_81,Huse_82,Baxter_83}, 
defined by Eq.~(\ref{def_H_Gas}) on the
square lattice with $K_{1N} \rightarrow -\infty$ and finite
$K_{2N}$, is known to have a tricritical point $(\mu_{tc},K_{2N,
tc})$ in the tricritical Ising universality class; the tricritical
point lies at $\mu_{tc}=-\ln [8(1+\sqrt{5})]$, $K_{2N, tc}=\ln
(3+\sqrt{5})$, with $\rho_{tc}=(5+\sqrt{5})/10$. Using the
transfer-matrix technique, Guo and coauthors\cite{Guo_02} determine
the critical point of the hard-core square lattice gas up to
the eleventh decimal place, $\mu_c=1.334 \, 015 \, 100 \, 277 \, 74
(1)$, $\rho_c=0.367 \, 742 \, 999 \, 041 \, 0 (3)$. Recently, Monte
Carlo simulations were carried out for square lattice gases with the
hard-core radius up to the fifth neighbors~\cite{Fernandes 07}. The
nature of phase transitions was found to be continuous for
exclusions up to 1N, to 2N, and to 4N, and to be discontinuous for
exclusions up to 3N and 5N, where symbols $iN$ represents the $i$th
neighbors.

\begin{figure}
\begin{center}
\includegraphics[scale=0.7]{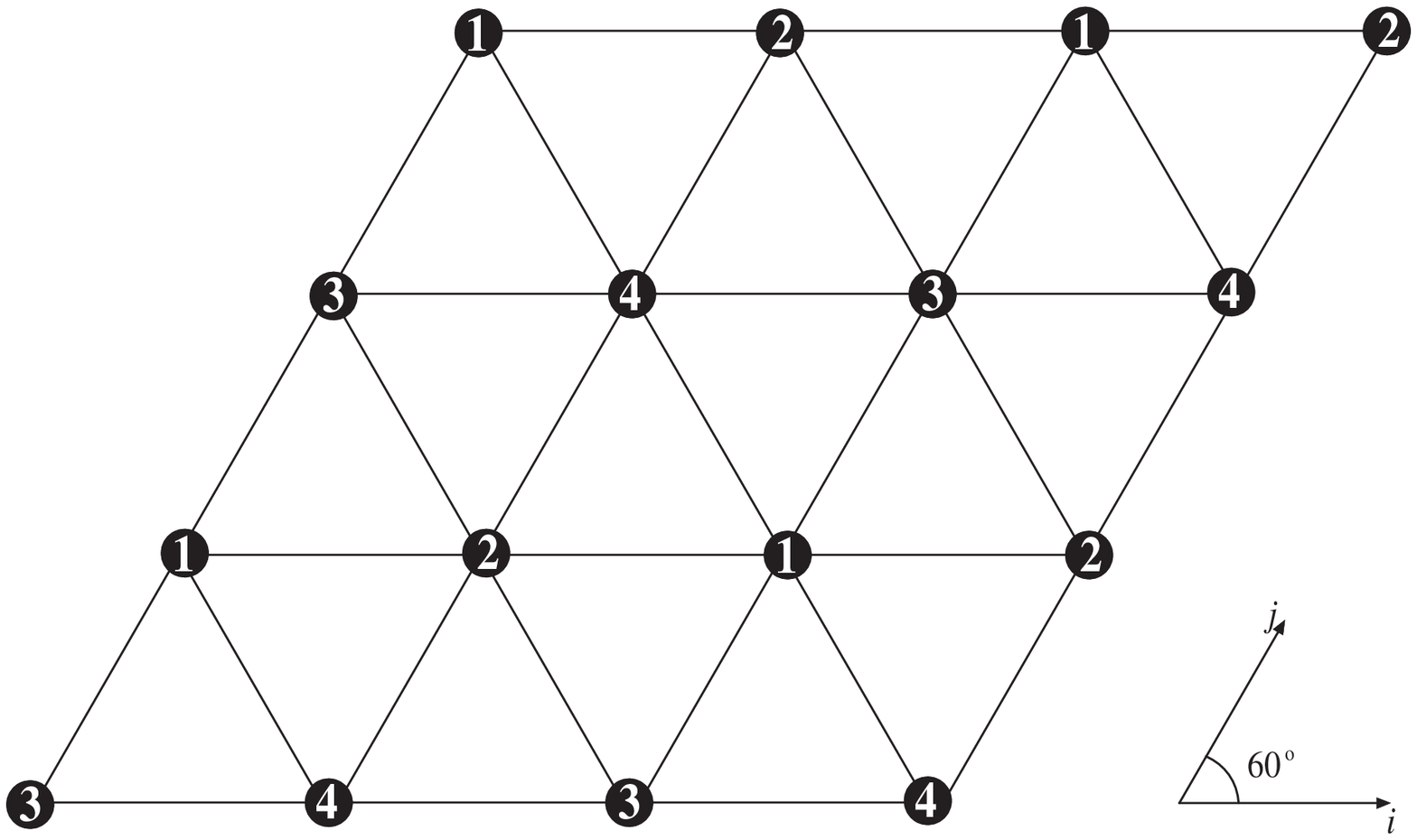}
\end{center}
\caption{The triangular lattice and its four sublattices. 
For chemical potential $\mu \rightarrow \infty$, one of the
sublattices is fully occupied.}
 \label{fig_lg4}
\end{figure}

In this work, we shall consider the hard-core lattice gas on the
triangular lattice with the first- and the second-neighbor
repulsions. The triangular lattice in this case can be divided into
four sublattices (see Fig.\ref{fig_lg4}), and for sufficiently high
density particles prefer to occupying one of the four
sublattices. Thus, one would expect that, if it is second order, the
melting of ordered phase should be in the 4-state Potts
universality class.
However, several studies at the end of 60s last century suggested
that the phase transition is first order~\cite{Orban_68,Runnels_71}. 
Later, Bartelt and
Einstein~\cite{Bartelt_84} reexamined this model by a
phenomenological renormalization--transfer-matrix scaling. The
largest system size in their study is $16 \times \infty$. Slowly
convergent finite-size corrections were observed.
It was estimated that the thermal and the magnetic critical exponents
are $y_t=1.400$ and $y_h =1.885$, respectively.
Despite the noticeable deviations from
the exact values $y_t=3/2$ and $y_h=15/8$~\cite{Nienhuis_87,Cardy_87},
these estimates are in favor of 
the 4-state Potts universality in view of the possible occurrence of
logarithmic corrections.

Here, we aim to provide an independent study of this model
by means of Monte Carlo simulations.
This seems justified since no rigorous argument exists
about the nature of the phase transition and
the evidence in Ref.~\cite{Bartelt_84} is not very strong. 
To properly analyze finite-size corrections, more
accurate numerical data, particularly for large system sizes, are
desirable~\footnote{Sometimes, when finite-site corrections are
not properly taken into account, wrong conclusions can be reached.
For instance, from the Metropolis simulations of the lattice-gas
model on the simple-cubic lattice with the first-neighbor repulsions,
Yamagata estimated the critical exponents as $\beta/\gamma=0.311 (8)$ and
$\gamma/\nu=2.38(2)$~\cite{Yamagata_96}, 
which would imply $y_h=2.689(8)$. This result
is significantly different from the general accepted value $y_h=2.4816
(2)$ for the Ising universality class in three dimensions~\cite{Deng_03}.}.
Our task becomes now feasible because of
the availability of efficient cluster algorithm for 
lattice-gas models--the geometric cluster algorithm--and the 
rapid development of
computer industry in the past few decades. 
The geometric cluster algorithm~\cite{Heringa_96,Heringa_97}
moves round a fraction of particles over the lattice 
according to geometric symmetries, such as the spatial inversion or rotation
symmetries of the triangular lattice; detailed
description will be given in Sec. II. The algorithm
does not change the total number of particles, and 
it is combined with the Metropolis steps in order to simulate
lattice-gas systems in the grand-canonical
ensemble. In comparison with simulations using the
Metropolis method only, critical slowing down is significantly
suppressed. Therefore, we are
able to simulate systems as large as $400 \times 400$ within
reasonable computer resources.

\section{Geometric cluster algorithm and sampled quantities}
\subsection{Geometric cluster algorithm}

The geometric cluster algorithm was first proposed by Dress 
and Krauth~\cite{Dress_95}
in the study of hard-core gases in continuous space. Unlike the well-known
Swendsen-Wang (SW) cluster method which flips spins, the elementary
operation in this algorithm is to move particles.
For hard disks, the percolation threshold of the cluster
formation process deviates
significantly from the phase transition of the model. This is unfortunate
since it affects the efficiency of the algorithm.

A single-cluster version of the geometric cluster algorithm was later
developed by Heringa and Bl\"ote \cite{Heringa_96,Heringa_97} for lattice
models like the Potts model and the lattice gases. 
Here, we shall briefly describe it in terms of
the lattice-gas model~(\ref{def_H_Gas}) with a soft-core 
radius of a lattice unit
(finite $K:\equiv K_{1N} < 0$ and all other couplings are zero) 
on the  square lattice with periodic
boundary conditions.  For such a geometry, one can set
a Cartesian coordinate by taking any two perpendicular
lines of lattice sites as the $x$ and the $y$ axis, respectively.
It can be seen that the Hamiltonian of the system is invariant
under geometric operations like reflections about the $x$ or the $y$
axis or inversion about the center of the coordinate. Further,
any configuration will be restored if an operation is subsequently applied
twice--namely, these operations are self-inverse. One can employ
any of such geometric operations to formulate a cluster algorithm.
Let a pair of nearest-neighboring sites $i,k$ be mapped onto
$i',k'$, respectively. One denotes the energy difference when a neighbor
$k$ of $i$ is interchanged with $k'$ as $\Delta_{ik}$, which is
$\Delta_{ik}=K(\sigma_i \sigma_k + \sigma_{i'} \sigma_{k'}-
\sigma_i \sigma_{k'} - \sigma_{i'} \sigma_{k}$).
The algorithm then involves the following steps:
\begin{enumerate}
\item Choose a random site $i$: both $i$ and $i'$ belong to the cluster.
\item Interchange $\sigma_i$ and $\sigma_{i'}$.
\item For all neighbors $k$ of $i$ that do not belong to the cluster yet,
do the following:
 \begin{itemize}
 \item If $\Delta_{ik} >0$, do the following with probability $p=1-\exp
      (-\Delta_{ik})$: (a) interchange $\sigma_k$ and $\sigma_{k'}$
      ($k$ and $k'$
      are included in the cluster), (b) write $k$ in a list of addresses (called
      the stack).
 \item If $\Delta_{ik}<0$, do nothing.
 \end{itemize}
\item Read an address $j$ from the stack. Substitute $j$ for $i$, and
  execute Step 3.
\item Erase $j$ from the stack.
\item Repeat Steps 4 and 5 until the stack is empty.
\end{enumerate}
When the stack is empty, the cluster is completed.
Since the elementary operation is to interchange
spins $\sigma_i$ and $\sigma_{i'}$, the total number of particles
does not change in the algorithm.

For the above geometric cluster steps, the detailed balance has
already been proved\cite{Heringa_96,Heringa_97}. The efficiency of this
algorithm for different models has also been demonstrated. For the
Ising model, it was shown that the percolation of the formed
clusters coincides with the thermal phase transition, reflecting
the efficiency of the algorithm. In fact, in the canonical ensemble
(the total number of particles is fixed), it can be shown 
that, for many models, no critical slowing down exist for some 
quantities~\cite{Deng_05}.

Here we shall formulate a full-cluster version of the geometric
cluster algorithm in an analogous way as the Edward-Sokal picture
for the well-known SW cluster method for the 
ferromagnetic Potts model~\cite{Swendsen_87,Edward_88}. We consider
the lattice-gas model~(\ref{def_H_Gas}) with finite
nearest-neighbor interactions ($K:\equiv K_{1N} <0$)
on a one-dimensional chain with sites labelled as
$i=\pm 1/2, \pm 3/2, \pm 5/2, \cdots $.
Instead of writing the Hamiltonian for a fixed number of particles
as a sum of the nearest-neighbor couplings like in Eq.~(\ref{def_H_Gas}),
we rewrite it as
\begin{equation}
{\mathcal H}|_{N_p =N}  = \sum_{i=1/2}^{\infty} {\mathcal H}_i =
- K \sum_{i=1/2}^{\infty} ( \sigma_i \sigma_{i+1} +
\sigma_{-i} \sigma_{-i-1}),
\label{def_H_Gas1}
\end{equation}
where $N_p=\sum_i \sigma_i$ and the constant
$N$ denotes the total number of particles.
The Hamiltonian~(\ref{def_H_Gas1}) is obtained by
applying the reflection about the center $i=0$--a geometric operation. 
In this form, the `building
blocks' of the Hamiltonian is no longer a pair of neighboring sites,
but two pairs of them.
If one only uses the spin-interchange operation $\eta :\equiv \sigma_i 
\leftrightarrow \sigma_{-i}$,
the energy associated with each building block 
is of two levels at most:
${\mathcal E}_1 (\vec{\sigma})
:\equiv -K(\sigma_i \sigma_{i+1} +\sigma_{-i} \sigma_{-i-1})$
and ${\mathcal E}_2 (\vec{\sigma})
:\equiv -K(\sigma_i \sigma_{-i-1} +\sigma_{-i} \sigma_{i+1})$.
The former ${\mathcal E}_1$ refers to the status that 
no operator $\eta$ is applied or it is applied 
at both vertices $i$ and $j$; instead,
the latter ${\mathcal E}_2$ means that $\eta$ is applied
at $i$ (or $j$) only.
The values of ${\mathcal E}_1$ and ${\mathcal E}_2$ depend on the
spin configuration $\vec{\sigma}$ on the building block. 
For the lattice-gas mode~(\ref{def_H_Gas1}),
these values are shown in Table~\ref{bond_weight}.
Let us denote
the lower and the upper level of ${\mathcal E}_1$ and ${\mathcal E}_2$ as
${\mathcal E}_{\rm low}$ and ${\mathcal E}_{\rm upp}$, respectively.
The statistical weight associated with each block
in Eq.~(\ref{def_H_Gas1}) reads
\begin{equation}
e^{-{\mathcal H}_i (\vec{\sigma})} = e^{-{\mathcal E}_{\rm upp}}
(1+v_i\delta_{{\mathcal E}_1, {\mathcal E}_{\rm low}}) \hspace{1cm}
\mbox{with} \hspace{2mm} (v_i=e^{{\mathcal E}_{\rm upp}-
{\mathcal E}_{\rm low}}-1) .
\label{exp_to_poly}
\end{equation}
On this basis, the partition sum becomes
\begin{equation}
{\mathcal Z}|_{N_p =N} = \sum_{\{\sigma\}: N_p=N} \; \;
\prod_{i=1/2}^{\infty} e^{-{\mathcal E}_{\rm upp} (\vec{\sigma})}
\prod_{i=1/2}^{\infty} (1+v_i \delta_{{\mathcal E}_1,
{\mathcal E}_{\rm low}}).
\label{def_Z_Gas}
\end{equation}
Analogously as mapping the Potts model onto the 
random-cluster model,
one introduces a bond variable $b_i$ to graphically represent the
expansion of the second product in Eq.~(\ref{def_Z_Gas}):
if the term $v_i \delta_{{\mathcal E}_1,{\mathcal E}_{\rm low}}$
is taken, an occupied bond $b_i=1$ is placed between sites $i$ and $i+1$;
otherwise, no bond is placed ($b_i=0$). This leads to a joint model
\begin{equation}
{\mathcal Z}|_{N_p=N} = \sum_{\{\sigma\}: N_p=N} \; \;
\prod_{i=1/2}^{\infty} e^{-{\mathcal E}_{\rm upp} (\vec{\sigma})}
\sum_{\{b\}} (v_i \delta_{{\mathcal E}_1, {\mathcal E}_{\rm low}})^{b_i},
\label{def_Z_Gas1}
\end{equation}
where the second sum is over all possible bond configurations that are
consistent with the spin configuration, and we have already assumed the
conventional symbol $0^0=1$. Given a spin
configuration $\{ \sigma \}$, the expression~(\ref{def_Z_Gas1}) allows
us to place bonds and construct clusters as follows: if the spin
configuration on a block $i$ is at the
lower-energy level ${\mathcal E}_{\rm low}$, one places a bond $b_i=1$
with probability $v_i/(1+v_i)$;
otherwise, place no bonds. Note that a bond connects four lattice sites,
since it is placed on the blocks.
Lattice sites connected through a chain of occupied bonds
form a cluster. The
condition $\delta_{{\mathcal E}_1, {\mathcal E}_{\rm low}}=1$ for a
block can be hold either by doing nothing or interchanging 
both spins ($\sigma_i
\leftrightarrow \sigma_{-i}, \sigma_{i+1} \leftrightarrow
\sigma_{-i-1}$). Thus, for a spin-jointed-bond configuration,
for each cluster one has the freedom to choose the do-nothing
or the spin-interchange operation,
and apply it to {\it all}
lattice sites within the cluster. Accordingly, a Swendsen-Wang-like
geometric cluster algorithm can be formulated as follows.
\begin{enumerate}
\item Choose a geometric transformation such that every building block in
the Hamiltonian consists of two pairs of neighboring couplings and the
associated energy is only of two levels under the spin-interchange
operation.
\item For each building block $i$ (containing four lattice sites), if
its spin configuration is at the lower-energy ${\mathcal E}_{\rm low}$,
place a bond with probability $p_i=v_i/(1+v_i)$; otherwise, place no bond.
\item Construct clusters according to the occupied bonds.
\item Independently for each cluster, randomly choose the do-nothing
or the spin-interchange operation with probability $1/2$; apply the
chosen operation to all lattice sites within the cluster.
\end{enumerate}
A Monte Carlo step is completed, and a
new spin configuration is obtained.  We expect that the analogy 
between our formulation of the geometric cluster
algorithm and the well-known SW method 
can help the reader to understand the 
geometric cluster algorithm.

We conclude this subsection by mentioning the following. (1), 
like the SW method, the essence
in the geometric cluster process is that the energy of 
each building block has two levels only under 
the spin-interchange operation (the energy of 
a building block may have more than two levels if other 
operations--like the spin-flip operation--are allowed).
(2), normally,
the rewriting of the Hamiltonian as a sum of proper building blocks
is obtained by applying some global geometric transformation, such as
the inversion about the center and reflection etc.
(3), in the canonical ensemble, if one uses 
the geometric cluster algorithm only, a large number of 
spatial transformations should be available such that
each lattice site in a system should be able to reach any other
lattice site by a finite number of geometric mappings. For the torus
geometry, this can be easily achieved since any site can serve as the
center of the aforementioned Cartesian coordinate.
In case that the geometric cluster method is itself non-ergodic,
it can be combined with other algorithms like the Kawasaki dynamic.
(4), for simulations in the grand-canonical ensemble, other Monte Carlo
methods have to be used.

\subsection{Sampled quantities}
For the lattice-gas model~(\ref{def_H_Gas}), the triangular lattice
is divided into four sublattices.
The particle density is then sampled as 
\begin{equation}
\rho^{(i)} =\frac{4}{V} \sum_{k \in {\mathcal T}^{(i)}} \sigma_k
\label{def_Ms}
\end{equation}
where $V=L \times L$ is the volume of the lattice and the 
sum is over each sublattice, labelled as $i=1,2,3,4$. The global particle
density is then $\rho=(\rho^{(1)}+\rho^{(2)}+
\rho^{(3)}+ \rho^{(4)})/4$. On this basis, we measured the
second and the fourth moment of the
magnetization density as
\begin{equation}
{\mathcal M} ^2=\frac{1}{3}
\sum_{i=1}^3 \sum_{j=i+1}^4 (\rho^{(i)}-\rho^{(j)})^2 \hspace{10mm}
\mbox{and} \hspace{10mm}
 {\mathcal M} ^4 = \left( {\mathcal M} ^2 \right)^2 \;,
\label{def_m2}
\end{equation}
where factor $1/3$ is for normalization purpose such that ${\mathcal M}^2$
is a unity when the chemical potential is infinite--one of the four sublattices
is fully occupied. The magnetic susceptibility is $\chi =V 
\langle {\mathcal M}^2 \rangle $.
Then, we define a dimensionless ratio as
\begin{equation}
Q=\frac{\<  {\mathcal M} ^2 \> ^2}{ {\< \mathcal M} ^4 \> } \, .
\label{def_Q}
\end{equation}
This ratio at criticality approaches a universal value for $L
\rightarrow \infty$, and is known to be very useful in 
estimating critical points.

Since a pair of first- (or second-) neighboring sites cannot
be both occupied by particles, we sampled the third-neighbor correlation
as an energy-like quantity
\begin{equation}
{\mathcal E} =\frac{1}{6V} \sum_{\{ij\} \in 3N} \sigma_i \sigma_k
\label{def_E}
\end{equation}
where the sum is over all the third-neighbor pairs. Correspondingly, a
specific-heat-like quantity is defined as ${\mathcal C}_e = V ( \<
{\mathcal E}^2 \>-\<{\mathcal E} \>^2 ) $. We also measured the
compressibility ${\mathcal C}_v = V( \< \rho^2 \> -\< \rho \>^2)$.

\section{Results}

Using a combination of the Metropolis and the geometric cluster
algorithm, we simulated the lattice-gas model on the triangular
lattice with first- and second-neighbor repulsions. Periodic
boundary conditions were used in the rhombus geometry shown in
Fig. \ref{fig_lg4}. System sizes took 15 values in range $8 \leq
L \leq 400$. Several geometric cluster steps are performed 
between subsequent Metropolis steps.
such that the total number of particles moved by the former is
also approximately equal to $V$. Significant critical slowing down 
was observed, with a dynamic exponent about $z \approx 1$.

Some primary simulations for relatively small system sizes were
first carried out to find the approximate location of the critical
point from the intersection of the $Q$ data for different
system sizes $L$ (we were also guided by the result $\mu_c \approx
1.7599$ in Ref.~\cite{Bartelt_84}). Then extensive simulations for large
sizes were performed near $\mu_c = 1.757$.

\begin{figure}
\begin{center}
\leavevmode
\includegraphics[scale=1]{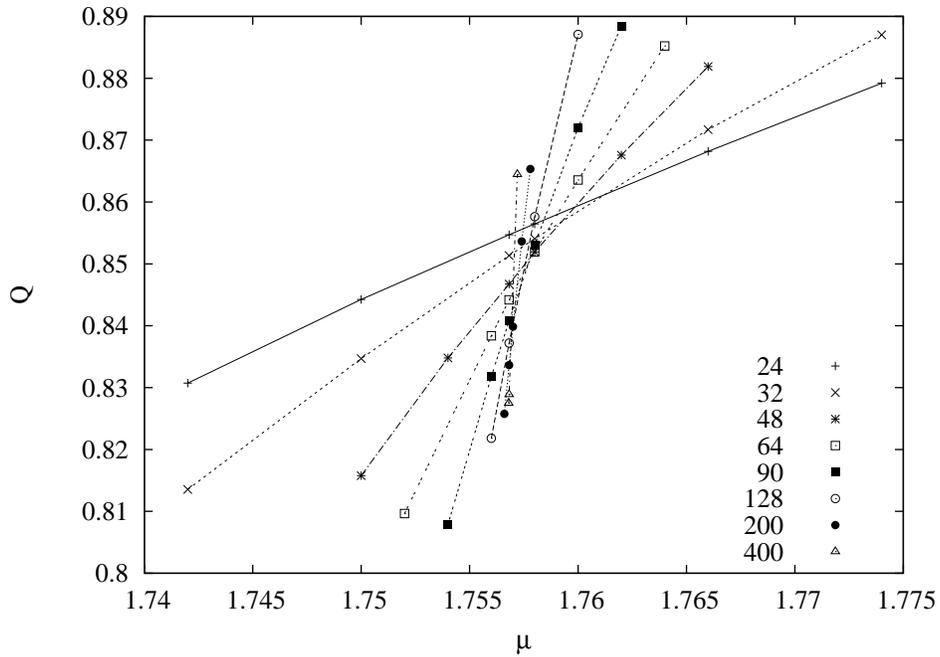}
\end{center}
\caption{Ratio $Q$ versus $\mu$ for various system 
sizes. The straight line segments, simply connecting the data 
points, are for illustration purpose. }
\label{fig_q}
\end{figure}

Figure \ref{fig_q} shows parts of the Monte Carlo data of the
dimensionless ratio $ Q$. 
According to the least-squares criterion, the $Q$ data were fitted 
by 
\begin{eqnarray}
Q(\mu, L)&=& Q_c+q_1(\mu-\mu_c)L^{y_t}+q_2(\mu-\mu_c)L^{2y_t} \nonumber\\
& &+ b_1L^{y_1}+b_2L^{y_2}+b_3L^{y_3}+r_1 L^{y_r}+c_1 (\mu -\mu_c)
L^{y_{t}+y_1} \; ,
\label{eq_q}
\end{eqnarray}
where $q_i$ , $b_i$ , $c_i$ , and $r_i$ are
unknown parameters, and $Q_c$ is the universal value. 
The terms with $q_i$ describe the contributions of the thermal field 
due to deviation from the critical point, those with exponent 
$y_i$ account for finite-size corrections, and the one with $c_1$ is for
the mixed effect of the relevant and irrelevant thermal fields. The
term with exponent $y_r=d-2 y_h$ arises 
from the regular part of the free energy, where the magnetic 
exponent $y_h$ was fixed at $15/8$ for the four-state Potts model.
The detailed derivation of the finite-size scaling formula~(\ref{eq_q})
can be found in Ref.~\cite{Deng_03}. In principle, one should include 
logarithmic corrections in Eq.(\ref{eq_q})~\cite{Salas_97}, 
since the transition is expected to be in the 4-state Potts 
universality class. Unfortunately, the limited system size 
does not allow us to include such corrections (accordingly, 
the statistical error 
margins of our following results should be taken carefully).
From the numerical results for the tricritical 4-state Potts 
model~\cite{Deng_05} where the marginal field is absent, we learn
that there exist correction terms with exponent $y_1=-1$. 
Thus, we set $y_1=-1$, $y_2=-2$, and $y_3=-3$.
Satisfactory fits can be obtained after a cutoff for small
system sizes $L  < 18$, which yield $\mu_c=1.75682(2)$,
$y_t=1.51(1)\approx 3/2$ and $Q_0=0.823(2)$. 
It is interesting to observe that, without logarithmic corrections, 
satisfactory fits can include data for rather small sizes and
the exponent $y_t=1.51 (1)$ agrees well with the exact value $3/2$.
This suggests that logarithmic corrections are small in the $Q$ data,
and thus the fitting results for $Q$ are more or less reliable.

We then fitted the $\chi$ data by 
\begin{eqnarray}
\chi &=& \chi_0 + L^{2y_h-2}[a_0+a_1(\mu-\mu_c)L^{y_t}+
a_2(\mu-\mu_c)L^{2y_t}\nonumber \\ 
&+& b_1L^{y_1}+b_2L^{y_2}+c_1(\mu-\mu_c)
L^{y_{t}+y_1}] \; , 
\label{eq_chi}
\end{eqnarray}
where $\chi_0$ stems from the regular part of the free energy, which
acts in Eq.~(\ref{eq_q}) as a correction term with 
exponent $y_r=2-2 y_h$.
The correction exponents were also set as $y_1=-1$ and $y_2=-2$.
After a cutoff for small systems $L < 20$,
we obtain $\mu_c=1.75683(1)$ , $y_t=1.489(9) \approx 3/2 $ and
$y_h=1.8748(8)\approx \frac{15}{8}$. The estimate of $\mu_c$
is consistent with that from $Q$, and those for $y_t$ 
and $y_h$ agree with the exact values.
If the exponent $y_t$ is fixed at $3/2$, one has
$\mu_c=1.75683(1)$ and $y_h=1.8743(7)$ after discarding the data
for $L < 18$.

The data for the particle density $\rho$ were fitted by 
\begin{eqnarray}
\rho&=&\rho_0+\rho_1(\mu-\mu_c)+ L^{y_t-d}[a_0+a_1(\mu-\mu_c)L^{y_t}+a_2
(\mu-\mu_c)L^{2y_t}\nonumber
\\&&+b_1L^{-1}+b_2L^{-2}+b_3L^{-3}] \; .
\label{eq_rho}
\end{eqnarray}
Satisfactory fits are obtained after a cutoff for small 
systems $L < 16$, and we have
$\mu_c=1.75682(2)$ , $y_t=1.440(5)$ and $\rho_c=0.180(4)$. 
The result $\mu_c=1.75680(3)$ agrees well with those obtained 
from magnetic quantities $Q$ and $\chi$. However, the value
$y_t=1.440(5)$ significantly differs from $3/2$.
This might imply that,
while additive logarithmic corrections are still small in 
energy-like quantities, multiplicative logarithmic corrections 
cannot be neglected.

The finite-size scaling formula of the specific-heat-like quantities
$C_e$ and $C_v$ reads
\begin{eqnarray}
C &=& r_0 +r_1 (\mu-\mu_c) +L^{2y_t-d}[a_0+a_1(\mu-\mu_c)L^{y_t}+
a_2(\mu-\mu_c)L^{2y_t}+a_3(\mu-\mu_c)L^{3y_t}\nonumber
\\&&+b_1L^{-1}+b_2L^{-2}+b_3L^{-3}+c_1 L^{y_{t}-1}(\mu-\mu_c) ] \; .
\label{eq_cv}
\end{eqnarray}
In the actual fits, the terms $r_0$, arising from the regular part 
of free energy, cannot be distinguished from the correction term $b_1 
L^{2y_t-d-1}=b_1 $, and so is for $r_1$ and $c_1$. Thus, 
we simply set $r_0$ and $r_1$ to be zero.
The data for $L \geq 18$ are well described by Eq.~(\ref{eq_cv}).
The fits for $C_v$ yield $\mu_c=1.75682(3)$ , $y_t=1.468(7)$ and
those for $C_e$ yield $\mu_c=1.75680(2)$ , $y_t=1.470(5)$. 
Again, the estimates of $\mu_c$ agree well with those from other 
quantities, while the values of $y_t$ differ significantly from $3/2$.

\begin{figure}
\begin{center}
\leavevmode
\includegraphics[scale=1]{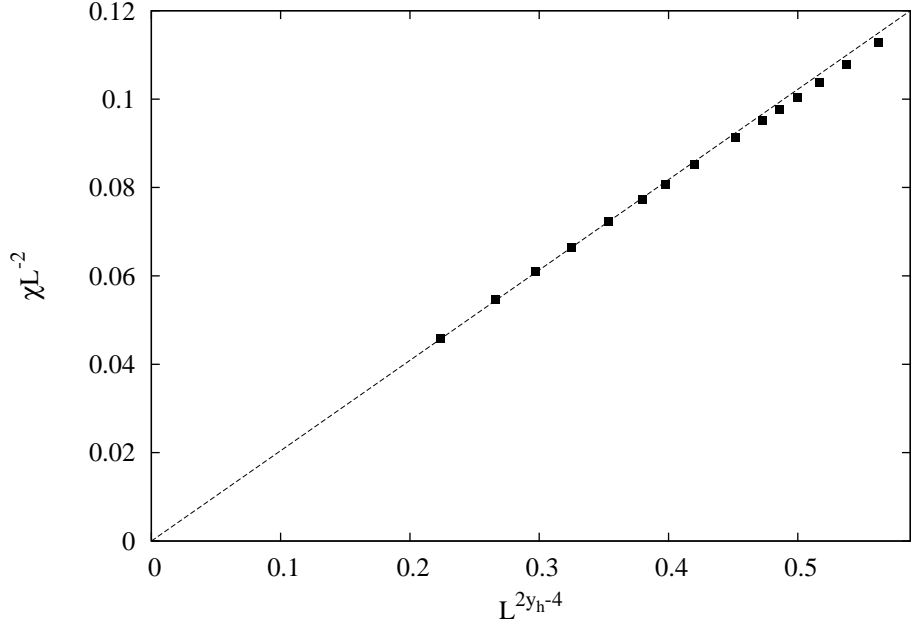}
\end{center}
\caption{Quantity $\chi/L^2$ at $\mu =1.756818 $
versus $L^{2y_h-4}$, with $y_h=15/8$. 
The statistical error bars are smaller
than the size of the data points. The dashed line is just for
illustration purpose.} \label{fig_chikc}
\end{figure}

\begin{figure}
\begin{center}
\leavevmode
\includegraphics[scale=1]{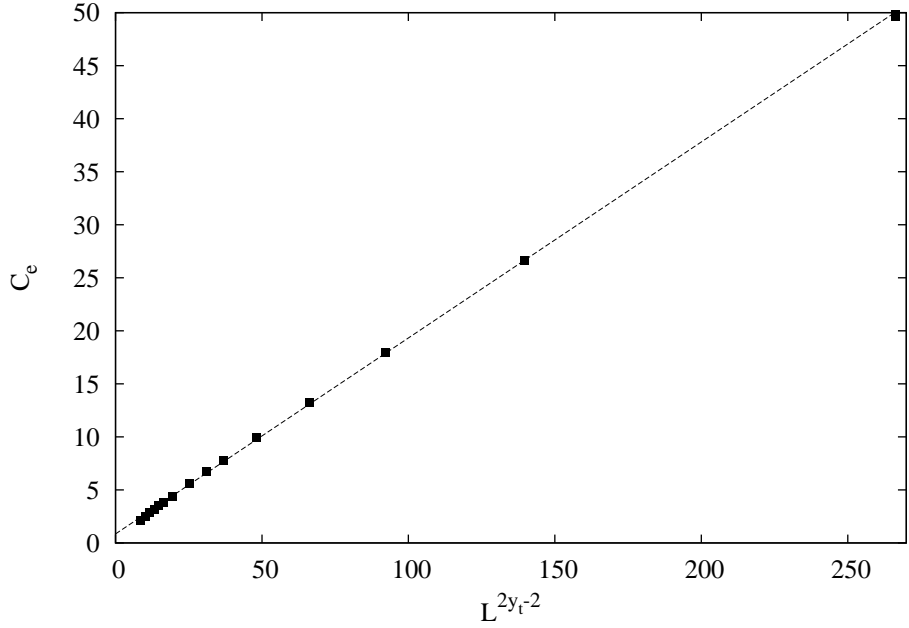}
\end{center}
\caption{Specific-heat-like quantity $C_e$ at $\mu =1.756818$
versus $L^{2y_t-2}$, where the value of $y_t=1.470$ is 
taken form the fit.  The statistical error bars of the data points
are in the same order of their size. The dashed line is for
illustration purpose.} \label{fig_cekc}
\end{figure}

We have simulations at $\mu=1.756818$, at criticality
within the estimated error bars. Thus, we could analyze various quantities 
right at the critical point.
The finite-size scaling behavior of $\chi$, $\rho$, and $C_e$ 
and $C_v$ at criticality is given by Eqs.~(\ref{eq_chi}), 
(\ref{eq_rho}), and (\ref{eq_cv}), respectively, by throwing out those 
$\mu$-dependent terms. The estimates of the associated critical 
exponents from these simplified analyses are consistent with those from the 
aforementioned fits. Figures \ref{fig_chikc} and \ref{fig_cekc} 
show the critical $\chi$ and $C_e$ data, respectively.


\section{Discussion}
In the language of the lattice gas systems, we formulate 
a Swendsen-Wang-like version of the geometric cluster algorithm that
has already found many applications~\cite{Deng_05,Liu_04}. 
Since our formulation is in line 
with the Swendsen-Wang algorithm for the ferromagnetic Potts model,
we expect that it will help the reader to understand the geometric
cluster method.

We then study the triangular lattice gases with the first- and the 
second-neighbor exclusion, using a
combination of the Metropolis and the geometric cluster algorithm.
The estimated critical point $\mu_c=1.75682(2)$ significantly 
improves over the existing result $1.7599$; to our knowledge, 
no report has been published yet for the critical particle 
density $\rho_c=0.180(4)$. 
The excellent agreement between the exact values and 
the numerical estimates $y_t =1.51 (1)$ and $y_h=1.8743 (7)$
give strong support for the expectation that the model is in the
4-state Potts universality class. On the other hand, the fitting
results from energy-like quantities imply that, although 
additive logarithmic corrections might be small, multiplicative
logarithmic corrections cannot be neglected at least in 
energy-like quantities. 

The fitting results are summarized in Table~\ref{tab_fit_result}.

{\bf Acknowledgements} This work was partially supported by the National 
Natural Science Foundation of China under Grant No. 10447111 and the Alexander 
von Humboldt Foundation of Germany (YD). 
One of us (YD) is greatly indebted
to Henk W.J. Bl\"ote, Timothy G. Garoni, and 
Alan D. Sokal for valuable discussions.

\begin{table}[ht]
\centering
\caption{\label{bond_weight} Energies ${\mathcal E}_1$ and ${\mathcal E}_2$ of
a building block in Eq.~(\ref{def_H_Gas1}). The associated bond weight 
for $K<0$ in the geometric cluster algorithm is also given. In 
the ``Example", the upper two sites are $i$ and $i+1$, and the lower 
are $-i$ and $-i-1$.  The filled (empty) circle represents the 
presence (absence) of a particle.} 
\begin{tabular}{c|c|c|c|c|c|c|c}
 Case        &  1                   &  2                    & 3                    & 4
             &  5                   &  6                    & 7   \\
\hline
             &  0 particle          &  1 particle           & \multicolumn{3}{c}{2 particles}
                                    &  3 particles          & 4 particles       \\
\hline
 Example     &  $\circ$--$\circ$    & $\bullet$--$\circ$    & $\bullet$--$\circ$   & $\bullet$--$\bullet$
             & $\bullet$--$\circ$   & $\bullet$--$\bullet$  & $\bullet$--$\bullet$  \\
             &  $\circ$--$\circ$    & $\circ$--$\circ$      & $\bullet$--$\circ$   & $\circ$--$\circ$
             & $\circ$--$\bullet$   & $\bullet$--$\circ$    & $\bullet$--$\bullet$  \\
\hline
${\mathcal E}_1$   & 0       &  0       &  0      &  $-K$    &  0      & $-K$     & $-2K$   \\
${\mathcal E}_2$   & 0       &  0       &  0      &   0      &  $-K$   & $-K$     & $-2K$   \\
\hline
$v$                & 0       &  0       &  0      &   0      & $e^{-K}-1$ & 0     & 0       \\
\end{tabular}
\end{table}

\begin{table}[ht]
\caption{Fitting results for various quantities. Symbol $L_{\rm min}$ 
is the minimum system size for which the Monte Carlo data are included
in the fit.}
\begin{tabular}{clllllll}
Quantity & $L_{min}$ & $\mu_c$  & $\rho_c$ & $y_t$     & $y_h$ \\
\hline
$Q$      & $20$ & $1.75682(2)$  &            & $1.51 (1)$  & \\
$\chi$   & $20$ & $1.75683(1)$  &            & $1.489(9)$  & $1.8748(8)$ \\
$\rho$   & $18$ & $1.75680(3)$  & $0.180(4)$ & $1.440(5)$  &             \\
${\mathcal E}$  
         & $12$ & $1.75680(2) $ &            & $1.48(2)$   &             \\
$C_e $   & $18$ & $1.75680(2)$  &            & $1.470(5)$  &             \\
$C_v $   & $18$ & $1.75682(3)$  &            & $1.468(7)$  &             \\
\hline
Previous &      & $1.7599$      &            & $1.400$     & $1.885$    
\end{tabular}
\label{tab_fit_result}
\end{table}
\end{document}